\documentclass[useAMS,usenatbib]{mnras}
\usepackage{times}
\usepackage{graphicx,latexsym, amssymb, amscd, psfrag}
\usepackage{epsfig,color}
\usepackage{epstopdf,subfig}
\usepackage{morefloats,rotating,float}
\usepackage{gensymb}
\usepackage{subfig,multirow,array}

\title[GAMA: Blue spheroids within 87 Mpc]{Galaxy And Mass Assembly (GAMA): Blue spheroids within 87 Mpc}
\author[Mahajan et al.]
{\parbox{\textwidth}{Smriti Mahajan$^{1},$\thanks{E-mail: \texttt{smritimahajan@iisermohali.ac.in}}
 Michael J. Drinkwater$^{2}$,  S. Driver$^{3,4}$, A. M. Hopkins$^{5}$, Alister W. Graham$^{6}$, S. Brough$^{7}$,  
 Michael J.I. Brown$^{8}$, B.W. Holwerda$^{9}$, Matt S. Owers$^{10,5}$, Kevin A. Pimbblet$^{11}$}\vspace{0.4cm} \\
 \parbox{\textwidth}{$^{1}$Indian Institute for Science Education and Research Mohali- IISERM, 
 Knowledge City, Manauli, 140306, Punjab, India \\ 
 $^{2}$School of Mathematics and Physics, University of Queensland, Brisbane, QLD 4072, Australia \\
 $^{3}$International Centre for Radio Astronomy Research (ICRAR), 
University of Western Australia, Crawley, WA 6009, Australia \\
 $^{4}$Scottish Universities' Physics Alliance (SUPA), School of Physics and Astronomy,
  University of St Andrews, North Haugh, St Andrews, KY16 9SS, UK \\  
 $^{5}$Australian Astronomical Observatory, PO Box 915, North Ryde, NSW 1670, Australia \\
 $^{6}$Centre for Astrophysics and Supercomputing, Swinburne University of Technology, Victoria 3122, Australia \\
 $^{7}$School of Physics, University of New South Wales, NSW 2052, Australia \\
 $^{8}$School of Physics and Astronomy, Monash University, Clayton, VIC 3800, Australia \\
 $^{9}$Department of Physics and Astronomy, 102 Natural Science Building, University of Louisville, Louisville KY 40292, USA \\
 $^{10}$Department of Physics and Astronomy, Macquarie University NSW 2109, Australia \\
 $^{11}$E.A.Milne Centre for Astrophysics, University of Hull, Cottingham Road, Kingston-upon-Hull, HU6 7RX, UK  }}

\newcommand{\s}{starburst}

\def\g{{GAMA }}
\def\s{{{\sc sigma }}}
\def\ser{{s\'{e}rsic }}

\def\mue{{$\langle \mu \rangle_e$}}
\def\re{{$R_{\rm{eff}}$}}
\def\ssf{{SFR/$M^*$}}
\def\bs{{BSph}}
\def\gr{{$(g-r)^0$}}
\def\smass{{$M^*/M_\odot$}}

\definecolor{grey}{rgb}{0.5,0.6,0.7}

\definecolor{amber}{rgb}{1.0,0.49,0.0}

\begin{document}

\date{}
\pubyear{2017}


\pagerange{\pageref{firstpage}--\pageref{lastpage}} 
\maketitle

\label{firstpage}
\begin{abstract}

 In this paper we test if nearby blue spheroid (\bs) galaxies may become the progenitors of star-forming spiral galaxies
 or passively-evolving elliptical galaxies. Our sample comprises 428 galaxies of various morphologies in the redshift range
 $0.002<z<0.02$ (8-87 Mpc) with panchromatic data from the Galaxy and Mass Assembly survey.
 We find that \bs~galaxies are structurally (mean effective surface brightness, effective radius) very similar to their
 passively-evolving red counterparts. However, their star-formation and other properties such as colour, age and metallicity
 are more like star-forming spirals than spheroids (ellipticals and lenticulars). We show that \bs~galaxies are statistically
 distinguishable from other spheroids as well as spirals in the multi-dimensional space mapped by  
 luminosity-weighted age, metallicity, dust mass and specific star formation rate. 
 
 We use HI data to reveal that some of the BSphs are (further) developing their disks, hence their blue colours.  
 They may eventually become spiral galaxies --- if sufficient gas accretion occurs --- or more likely fade into low-mass red
  galaxies.
\end{abstract}


 \section{Introduction}
 \label{intro}
  
 Traditionally, passively-evolving red galaxies have been associated with spheroidal morphology, while
 optically-blue, star-forming galaxies are found to be spiral or irregular. These observations are strengthened
 by the existence of two distinct locii for red and blue galaxies in the colour-magnitude space: the
 ``blue cloud" for the star-forming galaxies and the ``red sequence" for the passively evolving spheroids
 residing in dense environments. 
 
 In this paper we show that this colour-morphology relation breaks down for the blue spheroid (BSph) 
 galaxies\footnote{In the context of this paper, Blue spheroids are extremely
  blue and compact spheroid galaxies. Morphologically they resemble small elliptical galaxies or bulges of spirals.},
 especially below \gr~$\lesssim 0.5$, or \smass~$\lesssim 9$ where spheroidal galaxies are found to be
 predominantly blue and star-forming. Using observed and derived properties
 of galaxies, in this paper we examine the \bs~galaxies as the likely (i) progenitors of passively-evolving low-mass elliptical galaxies, 
 (ii) intermediate stage in the evolution of massive spirals or, (iii) unique population different from both the elliptical
 and spiral galaxies.
 
 Due to their fascinating nature \bs~galaxies have been the subject of many studies in the last few years. \citet{s09}
 for instance, presented a sample of \bs s selected by visual inspection from the Galaxy Zoo. Their sample comprised
 $L^*$ \bs~galaxies at $0.02<z<0.05$. They found that \bs~galaxies mostly reside in low-density environments, forming $\sim$ 6
 per cent of all spheroidal galaxies at low redshift. Using the deep Millennium Galaxy Catalogue 
 \citep[$\mu=26$ mag arcsec$^{-2}$;][]{liske03}, \citet{driver06} classified 11.2 per cent of the galaxies at $z\sim0$
 as \bs s, while \citet{cameron09} confirmed that \bs s are rare in the local Universe with a volume density of $(1.1 \pm 0.1)\times10^{-4}$
 h$^3_{70}$ Mpc$^{-3}$. 
 
 \citet{k09} found that \bs~galaxies become more common with decreasing
 stellar mass, such that their fraction increases from $\lesssim 2\% \sim 1-2 \times 10^{11} M_\odot$ to $\gtrsim 20\%$ 
 below $4-6 \times 10^9 M_\odot$. Since the latter mass scale coincides with the mass scale below which the mean 
 global atomic gas fraction increases for all types of galaxies \citep{k04,k08,janowiecki17}, Kanappan et al. suggested that the
 presence of cold gas may be crucial to the existence of \bs~galaxies. 
 Their findings were complimentary to that of \citet{noeske06} who analysed a sample of 26 luminous \bs~galaxies at high redshift
 ($\sim 0.2-1.3$) to show that most of the \bs s had a smaller, brighter star-forming component with an extended, 
 almost exponential disk-like component with scale length of
 $\lesssim 2$ kpc. Based on their results, \citet{noeske06} suggested that the majority of \bs~galaxies at high-$z$ will evolve into
 small disk galaxies or low-mass spheroids\footnote{Throughout this paper we collectively refer to lenticulars, \bs s and elliptical galaxies
 as spheroids unless stated otherwise.}. They neither find any evidence suggesting inside-out growth scenario which could 
 turn \bs s to large disks, nor do they see any disk growth around \bs s \citep[but see][for an alternative
  view]{graham15,graham16,graham17}. \citet{lopes16} also investigated a sample of low-$z$
 galaxies in various environments. Their analysis suggests that while highly-asymmetric \bs s may originate in mergers, the
 star formation histories of \bs s are likely to be heterogenous.
  
 In this paper we utilize the arsenal of data compiled by the Galaxy and Mass Assembly survey \citep[GAMA;][]{driver16}, which
 provides an unprecedented view of low-redshift galaxies using 21-band photometry and fibre-spectroscopic data as discussed
 in the following section. In Section~\ref{analysis} we analyse the physical properties of BSphs relative to other types of galaxies,
 and perform automatic classification of all galaxies in multi-dimensional parameter space in Section~\ref{s:kmeans}. We 
 analyse the neutral Hydrogen data for our sample, where available in Section~\ref{s:h1}. 
 We discuss our findings in the context of existing literature in Section~\ref{discuss}, and finally summarise our results in 
 Section~\ref{summary}. Throughout this paper we assume a $\Lambda$CDM concordance cosmological model with 
  $H_0=70$\,km s$^{-1}$ Mpc$^{-1}$, $\Omega_\Lambda=0.7$ and $\Omega_m=0.3$ to calculate all distances
  and magnitudes.

 \section{data}
\label{data}

 \subsection{Spectroscopic and photometric data}
 
 The GAMA survey is a combined spectroscopic and photometric multi-wavelength programme. GAMA has measured 
 photometry for over 230 deg$^2$ on the sky in 21 wavebands, and obtained spectroscopic redshifts for 
 $\sim 300,000$ galaxies ($z\lesssim0.25$) \citep{baldry10,robotham10,driver11,hopkins13,driver16}.   
 
  The sample used in this paper is the same as that used in \citet[][hereafter paper I]{mahajan15}, 
  but with some of the derived parameters updated in accordance with the latest versions of the GAMA II catalogues \citep{liske15}. 
  We direct the reader to Paper I for details of sample selection and its characteristics, highlighting only the most
  relevant aspects below briefly for completeness. Our sample comprises 428 galaxies\footnote{All but four galaxies from Paper I
  are excluded because of missing data in GAMA II catalogues.} selected to have very well constrained
  spectroscopic redshifts in the range $0.002< z_{TONRY}<0.02$\footnote{$z_{TONRY}$ uses the flow model described in \citet{tonry00}.}.
  This redshift range was chosen to exclude
  galactic stars and enable morphological classification of galaxies by visual inspection using shallow ($ 53.9$ sec per band)
  Sloan Digital Sky Survey (SDSS; data release 7) imaging. The spectroscopic campaign for GAMA is based on the SDSS imaging
  complete to $r<19.8$ mag, which translates to $M_r=-14.9$ mag at the maximum redshift ($z=0.02$) of data used in this paper. 
  
 We use the matched-aperture photometry measured across 21 wavebands for GAMA galaxies using the Lambda Adaptive
 Multi-Band Deblending Algorithm in R \citep[{\sc LAMBDAR};][]{wright16}. Specifically, we use LambdarSDSSgv01, 
 LambdarSDSSrv01 and LambdarInputCatUVOptNIRv01 catalogues from the LAMBDAR data management unit (DMU). 
 The magnitudes are then $k$-corrected to $z=0$ using the $k$-corrections from the kcorr\_auto\_z00v05 DMU
 \citep{loveday12}. We note that the $k$-corrections in this DMU have been obtained using the SDSS model magnitudes
 and GAMA matched aperture AUTO magnitudes from the ApMatchedCatv06 DMU. Since the correlation between
 the AUTO and {\sc LAMBDAR} magnitudes for our sample is better than 95 per cent\footnote{$>98$ per cent excluding irregular galaxies
 from our sample.} they can be applied to {\sc LAMBDAR} magnitudes for our sample. The mean $k$-correction for our sample in 
 the $g$ and $r$-bands is $\sim0.009$ mag.

 \begin{table*}
\caption{Complete sample of 428 galaxies (A complete version of this table is available online). The morphology classification (column 4) is:  
1: Elliptical, 2: Spiral, 3: Irregular, 5: BSph, 6: LSB and 11: Lenticular.}
\begin{tabular}{ccccccccc}
\hline 
GAMA ID & $M_r$ & $\Delta M_r$ & Morphology & z &  ALFALFA ID & HI Flux ($S_i$) & $\Delta S_i$ & log HI \\ 
              & mag  & mag &   &  &  & Jy km s$^{-1}$ & Jy km s$^{-1}$ & $M_{\odot}$  \\  \hline \hline
 105589  &  19.67  &    0.03   &    3      &    0.019   & - & - & - & -  \\
  106916  &  19.58  &    0.04    &   6    &      0.015   & 249428  & 0.63 &    0.05 &    8.72  \\
  107137   &    13.94   &   0.01    &   11    &     0.015   & - & - & - & -  \\
  107226   &    19.39  &    0.03   &    6   &       0.016   & - & - & - & - \\ 
  116572   &    19.84   &   0.02   &    6   &       0.020  &  - & - & - & - \\   
  117059   &    19.87  &    0.04    &    5   &       0.011   & - & - &  - &  -  \\
  118764    &   18.47 &    0.09   &     3     &     0.011    & - & - & - & - \\ 
  119004   &    17.87   &    0.05   &      3     &     0.013  & - & - & - & -  \\ \hline
\end{tabular}
\label{data}
\end{table*}

 \begin{figure}
 \centering{
 {\rotatebox{270}{\epsfig{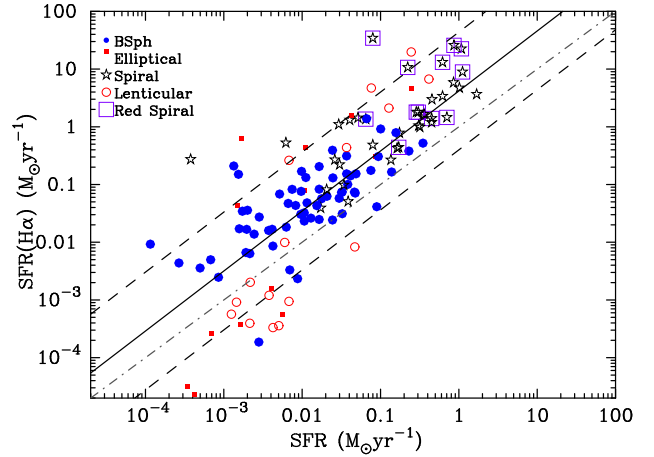}}}}
 \caption{Comparison between {\sc magphys}-derived ``average" SFR and the instantaneous SFR
 measured from the H$\alpha$ emission line. The solid and the dashed {\it black} lines represent the linear least square
 fit and $\pm 1\sigma$ deviation therein to the 415 galaxies for which data are available. The {\it grey} dot-dashed line 
 represents equal SFR on both axes. 
 For our sample, a constant offset can be applied to the {\sc magphys}-derived SFR to get instantaneous SFR.  }
 \label{sfr-compare}
 \end{figure}

  \subsection{Physical properties of galaxies}
  
 The best-fit values for several physical parameters for all galaxies in the three equatorial regions of the
 GAMA survey have been obtained by running the spectral energy distribution fitting code Multi-wavelength
  Analysis of Galaxy Physical Properties \citep[{\sc magphys};][]{dacunha08} on the 21-band photometry taken 
  from the Lambdarcatv01 DMU \citep{driver16}.    
  We use the stellar mass ($M^*$), star formation rate (SFR), $r$-band light-weighted age and metallicity ($Z$) 
  obtained from the Magphys DMU (version 6) in our analysis below. 
  
 The SFR derived from {\sc magphys} is an integrated measure of the SFR and therefore represents the star formation
 activity of a galaxy averaged over a long period of time (0.1 Gyr for {\sc magphys}). It may thus be argued that for galaxies undergoing a strong
 burst of star formation, {\sc magphys}-derived SFR may be very different from the instantaneous SFR. In order to 
 test this hypothesis we converted the H$\alpha$ equivalent width\footnote{The EW(H$\alpha$) are obtained from the 
 SpecLineSFRv05 \citep{gordon17}.} to luminosity using eqn. 5 of \citet{hopkins03}\footnote{We 
 assume a constant stellar absorption correction of 2.5 \AA~as in \citet{gunawardhana13}.}. The instantaneous ($< 10$ Myr)
 SFR is then estimated using the luminosity-to-SFR conversion factor given by \citet{kennicutt98}.

 Figure~\ref{sfr-compare} shows a comparison between the two measures of  SFR  for our sample. On average, the
 {\sc magphys}-derived SFR is $\sim 0.6$ dex lower than the H$\alpha$ SFR. We therefore conclude that for the sample used here, 
 the average SFR can be converted to an instantaneous SFR by using a simple scaling factor. The use of the latter instead of 
 former will only change our results qualitatively. We therefore use the {\sc magphys}-derived ``average" SFR throughout.
     
 \begin{figure}
 \centering{
{\rotatebox{270}{\epsfig{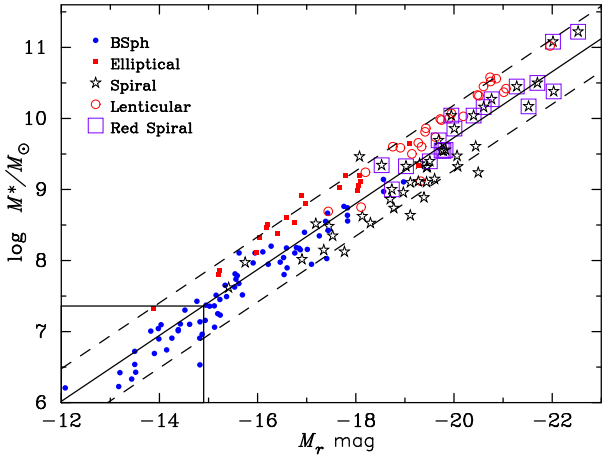}}}}
\caption{Stellar mass (${\rm log}M^*/M_{\odot}$) as a function of $r$-band absolute magnitude for our sample. The 
rectangular region at the bottom left marks the zone of incompleteness in this space.}
\label{mag-mass}
\end{figure}
  
  Figure~\ref{mag-mass} shows that the absolute $r$-band magnitude and {\sc magphys}-derived $M^*$ for our sample are
  well correlated\footnote{Although not shown for clarity, all 428 galaxies were included in evaluating the least square fit
  relation and the scatter therein.} as ${\rm log}~M^*/M_{\odot}=0.450-0.464M_r \pm 0.458$. Based on this figure and the
  limiting magnitude of $M_r=-14.9$ mag, in the following we adopt ${\rm log}~M^*/M_{\odot}=7.36$ as the
  limiting stellar mass for our sample.  
  16.3 per cent (27/165) of the galaxies shown in Figure~\ref{mag-mass} fall in the incomplete zone (the fraction 
  increases to $\sim 30$ per cent when Irr and LSB galaxies are included). 
  For completeness we show all data points in the following figures, highlighting the ones in the
  incomplete zone of the $M^*$-$M_r$ space.

 \subsection{Morphological classification}
  
  \begin{figure*}
  \centering
  \begin{tabular}{cccc}
\subfloat{\rotatebox{270}{\epsfig{file=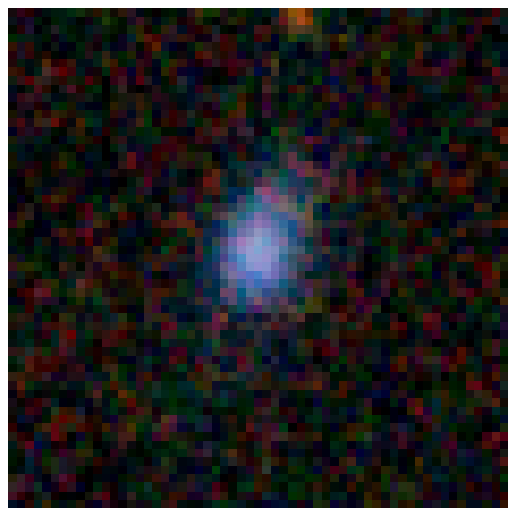,width=4cm}}} &
\subfloat{\rotatebox{270}{\epsfig{file=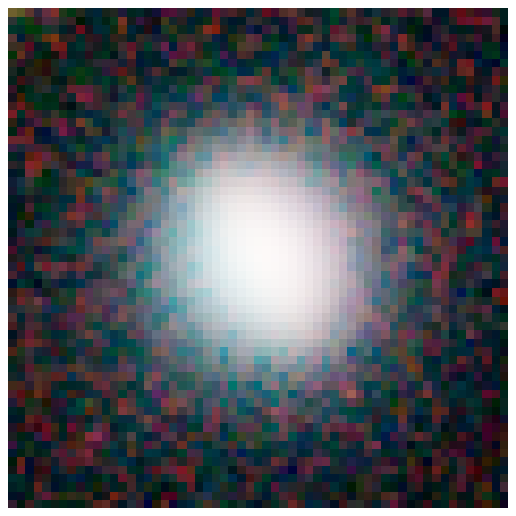,width=4cm}}} &
\subfloat{\rotatebox{270}{\epsfig{file=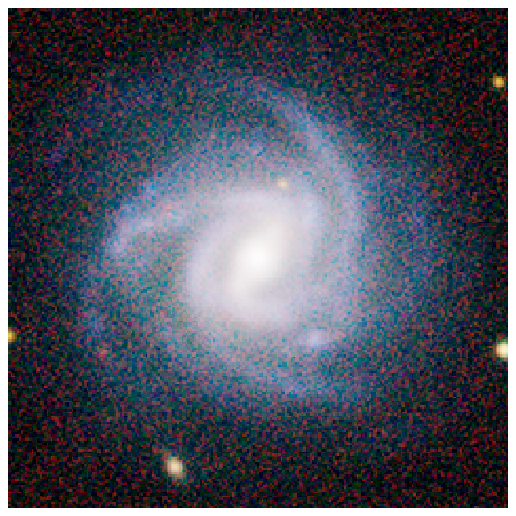,width=4cm}}} &
\subfloat{\rotatebox{270}{\epsfig{file=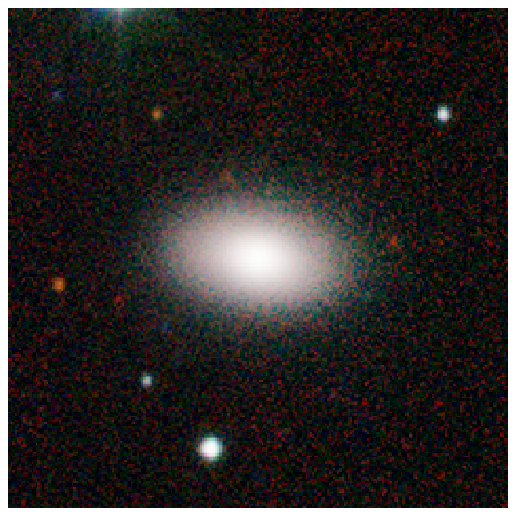,width=4cm}}} \\
\subfloat{\rotatebox{270}{\epsfig{file=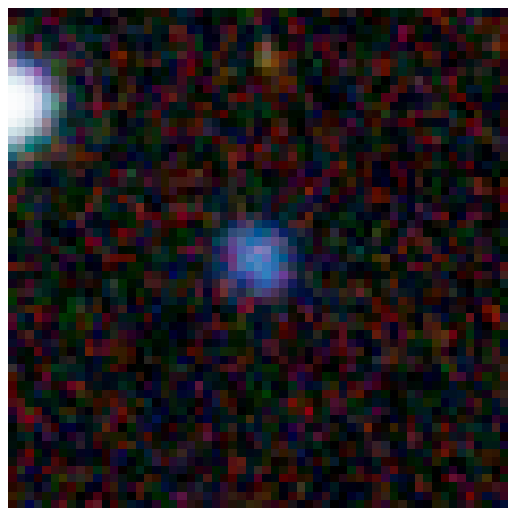,width=4cm}}} &
\subfloat{\rotatebox{270}{\epsfig{file=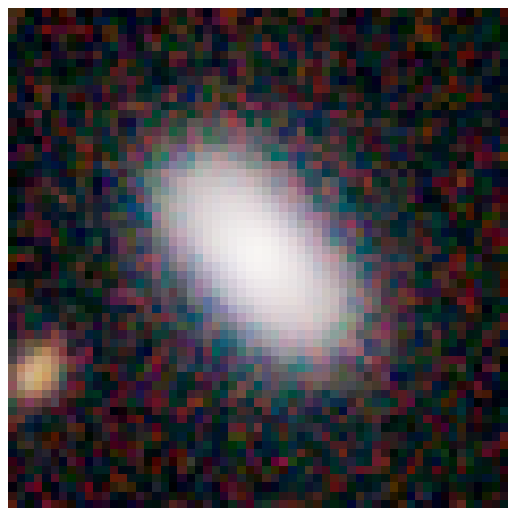,width=4cm}}} &
\subfloat{\rotatebox{270}{\epsfig{file=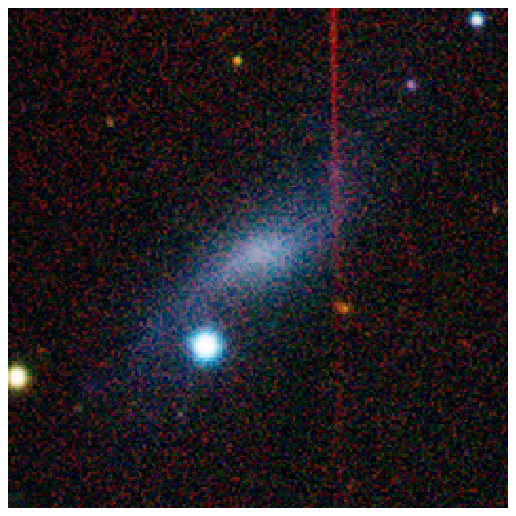,width=4cm}}} &
\subfloat{\rotatebox{270}{\epsfig{file=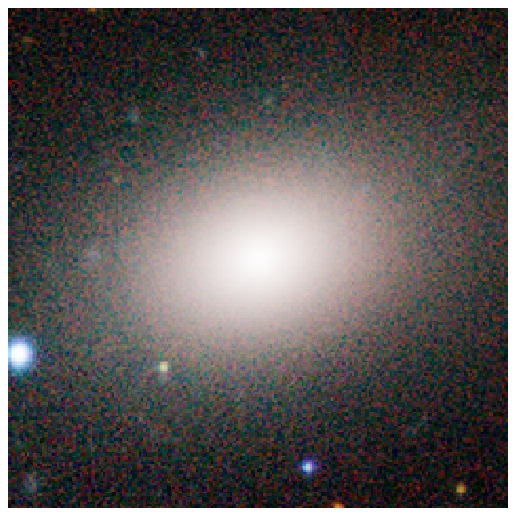,width=4cm}}} \\
\subfloat{\rotatebox{270}{\epsfig{file=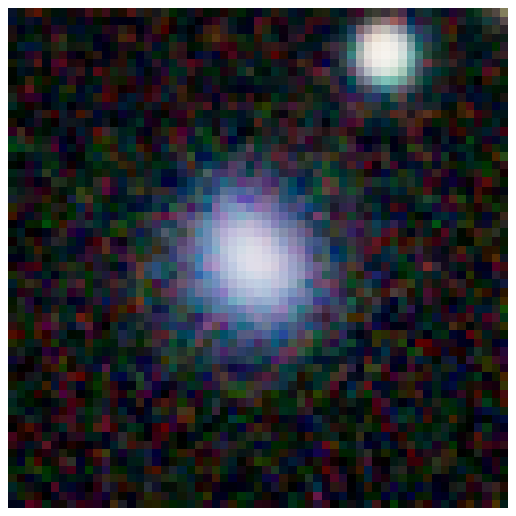,width=4cm}}} &
\subfloat{\rotatebox{270}{\epsfig{file=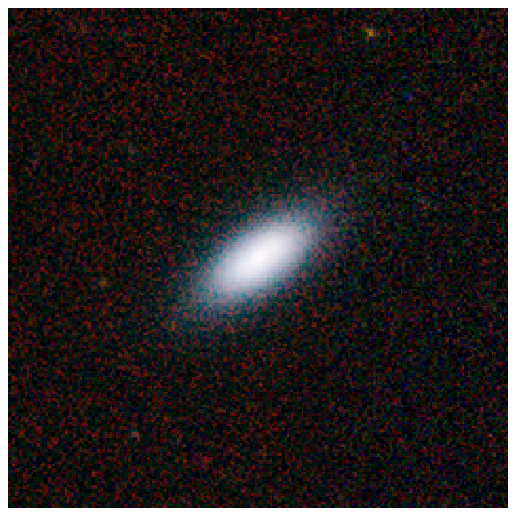,width=4cm}}} &
\subfloat{\rotatebox{270}{\epsfig{file=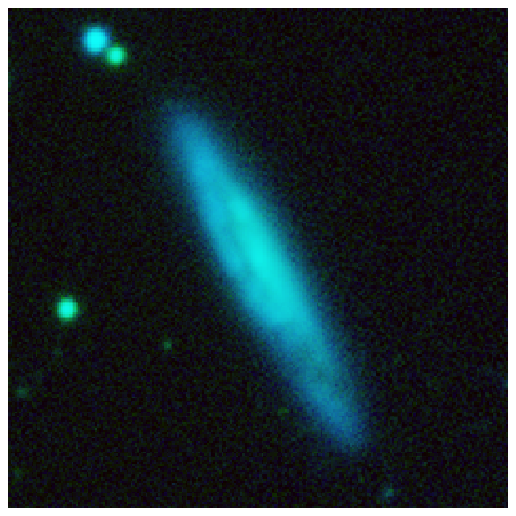,width=4cm}}} &
\subfloat{\rotatebox{270}{\epsfig{file=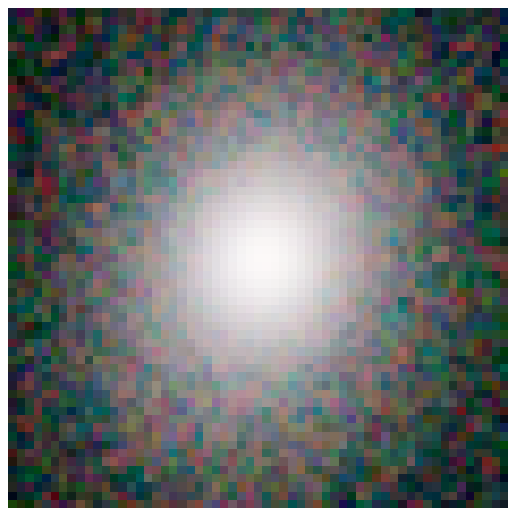,width=4cm}}} \\
\end{tabular}
\caption{Some examples of different morphological types of galaxies. Images are grouped along columns ({\it from left}): (i) Blue spheroids,
 (ii) Ellipticals, (iii) Spirals, and (iv) Lenticulars. Each $giH$ image is scaled to a constant physical size of $20 \times 20$ square kpc at
 the redshift of the galaxy.  }
 \label{images}
  \end{figure*}
  
 The morphological classification of galaxies used in this work is the same as that in paper I and we direct the reader to
 section 3.1 therein for details, briefly summarising our methodology below for completeness. 
 Visual classification of the three-colour $giH$ images\footnote{Each image is generated using the SDSS $g,i$ and the UK
 Infrared Telescope (UKIRT) Infrared Deep Sky Survey (UKIDSS) Large Area Survey (LAS) $H$-band data. 
 http://www.ukidss.org/surveys/las/las.html} of all galaxies in the redshift range
 $0.002$-$0.02$ was done by Mahajan, Driver and Drinkwater multiple times. Some representative examples of the postage stamp
 images classified into different categories are shown in Figure~\ref{images}. We found that in the chosen redshift range
 our data could be categorised into six classes: 
 \begin{itemize}
\item {\it Elliptical:} Galaxies which are morphologically elliptical in shape. They are mostly red in colour.  
 \item {\it Spirals:} Galaxies showing well-defined spiral arms or clearly identifiable edge-on disks. These galaxies
 often show conspicuous signs of ongoing star formation, such as HII regions, and stellar associations forming spiral arms.
 \item{\it Lenticulars:} Red, disk galaxies with a resolved nucleus. These galaxies are mostly big and bright, occasionally showing
 signs of some ongoing star formation in rings around nucleus, or low surface brightness disks without spiral arms.
 \item {\it Blue spheroids (BSph):} Colour plays a key role in successfully identifying these galaxies. They are very blue
 and generally compact spheroids, morphologically similar to small elliptical galaxies or bulges of spiral galaxies.
 \item{\it Low surface brightness (LSB) galaxies:} These extended objects show very poor contrast with the
 background in the five-band SDSS imaging. We note that a substantial fraction of these galaxies may have been
 misclassified due to the very shallow imaging data used here. Many of these galaxies may also be classified as
 irregular, and as we will show below, these two classes overlap in most of the parameter space explored here.  
 \item {\it Irregulars (Irr):} All confirmed extended sources that do not belong to any of the above categories.
\end{itemize}
 Although we did not use luminosity in our classification scheme, the results were luminosity-dependant, such that the latter
  three classes (BSphs, LSBs and Irr) dominate the low-luminosity regime in our sample, with irregulars being the most dominant
  population ($\sim 45$ per cent), followed by BSphs ($\sim 17$ per cent).
  
 415/428 ($97$ per cent) of our sources are also found in the VisualMorphologyv03 DMU of GAMA. In our chosen
 redshift range we find 57 ($14$ per cent) galaxies are classified as the ``little blue spheroids" in the 
 VisualMorphologyv03 DMU, of which 29 ($40$ per cent) are also classified as \bs s in this paper. 
    
 Table~\ref{data} gives the GAMA II IDs, $r$-band magnitude and the uncertainty in it, and
 redshifts compiled from the above mentioned DMUs along
 with the morphological classification (as per paper I) for all the galaxies in our sample. The morphological classification 
 is 1: Elliptical, 2: Spiral, 3: Irregular, 5: BSph, 6: LSB and 11: Lenticular. Since the focus of this paper is to compare \bs s with spirals
 and other spheroids, in the following we only show the 165 galaxies which are identified into one of these morphology 
 classes unless specified otherwise. A complete version of Table~\ref{data} is available online. Any further information
 for any of the 428 galaxies from this sample can be obtained from the GAMA
 website\footnote{http://www.gama-survey.org/} by using the unique GAMA IDs. 

\section{Analysis of physical properties}
 \label{analysis}
 
 In this section we discuss the trends in various physical properties for the galaxies in our sample, briefly discussing how each of
 them may contribute towards our understanding of the evolution of the \bs~galaxies.
 
 \subsection{Luminosity and Colour}
 
 \begin{figure}
 \centering{
{\rotatebox{270}{\epsfig{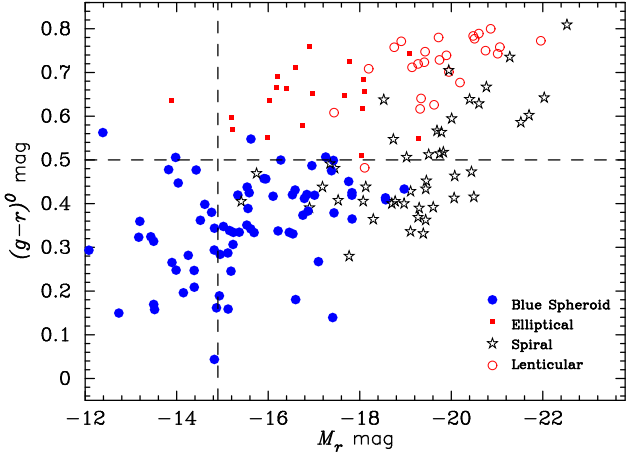}}}}
\caption{The spiral and spheroidal galaxies (Lenticulars, Ellipticals and BSphs) in the colour-magnitude space. 
 The horizontal dashed line is the colour-cut used for separating blue and red spiral galaxies at $(g-r)^0=0.5$ mag, and
 the vertical dashed line is the $r$-band completeness limit corresponding to $z_{max}=0.02$. }
\label{col-mag}
\end{figure}
 
 In Figure~\ref{col-mag} we show the distribution of spheroidal and spiral 
 galaxies in the colour-magnitude diagram. While the BSphs occupy the region with the bluest colours irrespective
 of the magnitude, the lenticulars are the most luminous and optically red in colour. The elliptical galaxies have
 red colour similar to lenticulars, but
 are less luminous than the latter. Almost 2/3rd of the spiral galaxies have $(g-r)^0 \sim 0.4\pm0.1$ mag, but 
 the rest of them acquire increasingly redder colour as they become more luminous than $M_r \sim -18.5$ mag. 
 Figure~\ref{col-mag} as well as visual inspection of spiral galaxies in our sample suggests that $(g-r)^0=0.5$ mag 
 is a good divider for segregating blue spirals from red ones. The red spirals are marked in all the following figures
 to distinguish them from their blue counterparts. 
 
 Figure~\ref{col-mag} shows that the passively-evolving galaxies (ellipticals and lenticulars) and star-forming spirals
 and BSphs form a continuous distribution in the optical colour-magnitude space.    

\subsection{Star formation}
\label{s:sf}

  \begin{figure}
 \centering{
 {\rotatebox{270}{\epsfig{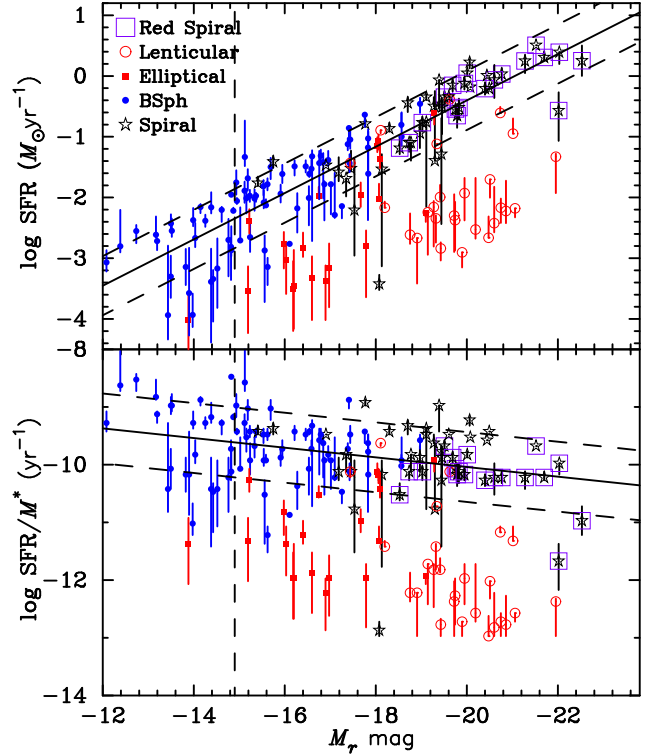}}}}
 \caption{The SFR ({\it top}) and SFR/$M^*$ ({\it bottom}) for our sample of galaxies
 as a function of $M_r$. Symbols and vertical line are same as in Fig.~\ref{col-mag}. The solid line is the
 linear least square fit to the BSphs and spiral galaxies, while the dashed lines represent 1-$\sigma$ deviation
 in the fitted relation. }
 \label{sfr}
 \end{figure}
 
 In Figure~\ref{sfr} we show the distribution of the SFR and \ssf~for galaxies with different morphologies as a function
 of their $r$-band absolute magnitude. The BSphs and spirals form a sequence such that the SFR increases 
 with luminosity, according to the fitted relation represented by a solid line in the top panel of Figure~\ref{sfr}.
 The passively evolving ellipticals and lenticular galaxies also seem to follow a similar relation although with a lower
 normalisation factor. This trend is replicated for the \ssf, although with a lower slope for the fitted relation since
 most of the star-forming galaxies (BSphs and spirals) have \ssf~in the range $10^{-8}$ to $10^{-10}$ yr$^{-1}$, unlike their log SFR
 which varies by $\sim 5$ dex. The log \ssf~for the passively evolving ellipticals and lenticulars
 on average decreases by $> 2$ dex relative to the star-forming galaxies. The $r$-band
 absolute magnitude is related to the SFR and \ssf~in Figure~\ref{sfr} by the relations: 
 \begin{equation}
\rm log\, SFR (M_{\odot}yr^{-1})=-0.381{\it M_{r}}-8.029 \pm 0.486
\end{equation}
and
\begin{equation}
\rm log\, SFR/M^{*} (yr^{-1})=0.084{\it M_{r}}-8.367 \pm 0.605
\end{equation}
where, the uncertainty is the 1-$\sigma$ deviation in the fitted linear least squares relation for star-forming galaxies
 (\bs s and spirals only). 
 
 These trends in the SFR and \ssf~show that \bs~galaxies are closer to spirals than ellipticals
 in the sense that they follow the same relation as the spirals, mostly occupying the lower (higher) edge of the SFR (\ssf) distribution.

\subsection{Age and Metallicity}
 
 \begin{figure}
 \centering{
 {\rotatebox{270}{\epsfig{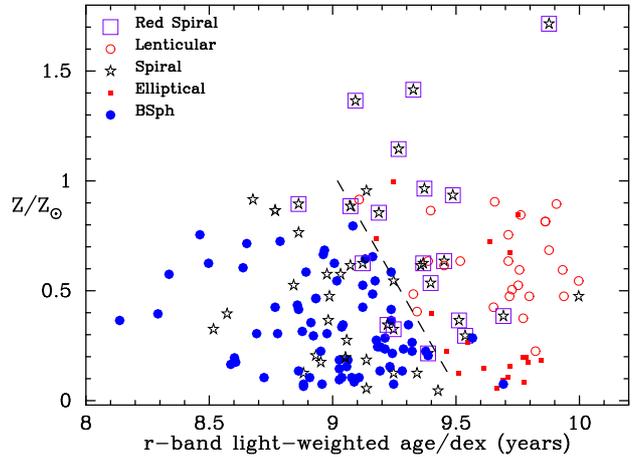}}}}
 \caption{Galaxies in the age-metallicity plane. The {\it purple open squares around stars} represent red spiral galaxies identified in 
 Figure~\ref{col-mag}. The dashed line is shown to guide the eye to the division between star-forming and passive
  galaxies in this space. }
 \label{age-z}
 \end{figure}

 In Figure~\ref{age-z} we show the distribution of our galaxies in the age-metallicity space. 
 The star-forming spirals and \bs s cluster together in the left of the diagram at lower age and metallicity ($Z$), while the ellipticals
 coincide with the lenticulars in the other half. As expected, most of the red spirals 
 identified in Figure~\ref{col-mag} are found closer to the dividing line, or coincide with the passively-evolving
 galaxies in this age-metallicity space. At least half of the red spirals on average also have greater $Z$ than the
 lenticulars. We show an approximate dividing line in Figure~\ref{age-z} to guide the eye to the suggested division between 
 the star-forming and passively-evolving galaxies. The significance of this line will become clear in the next section
 where we employ an automated algorithm to find clusters of galaxies based on various combinations of other physical properties.  
 
 Since \bs~galaxies coincide with star-forming spirals in Figure~\ref{age-z}, it can be concluded that they may have
 shared similar star formation histories.   
 
 \begin{figure*}
\centering{
{\rotatebox{0}{\epsfig{file=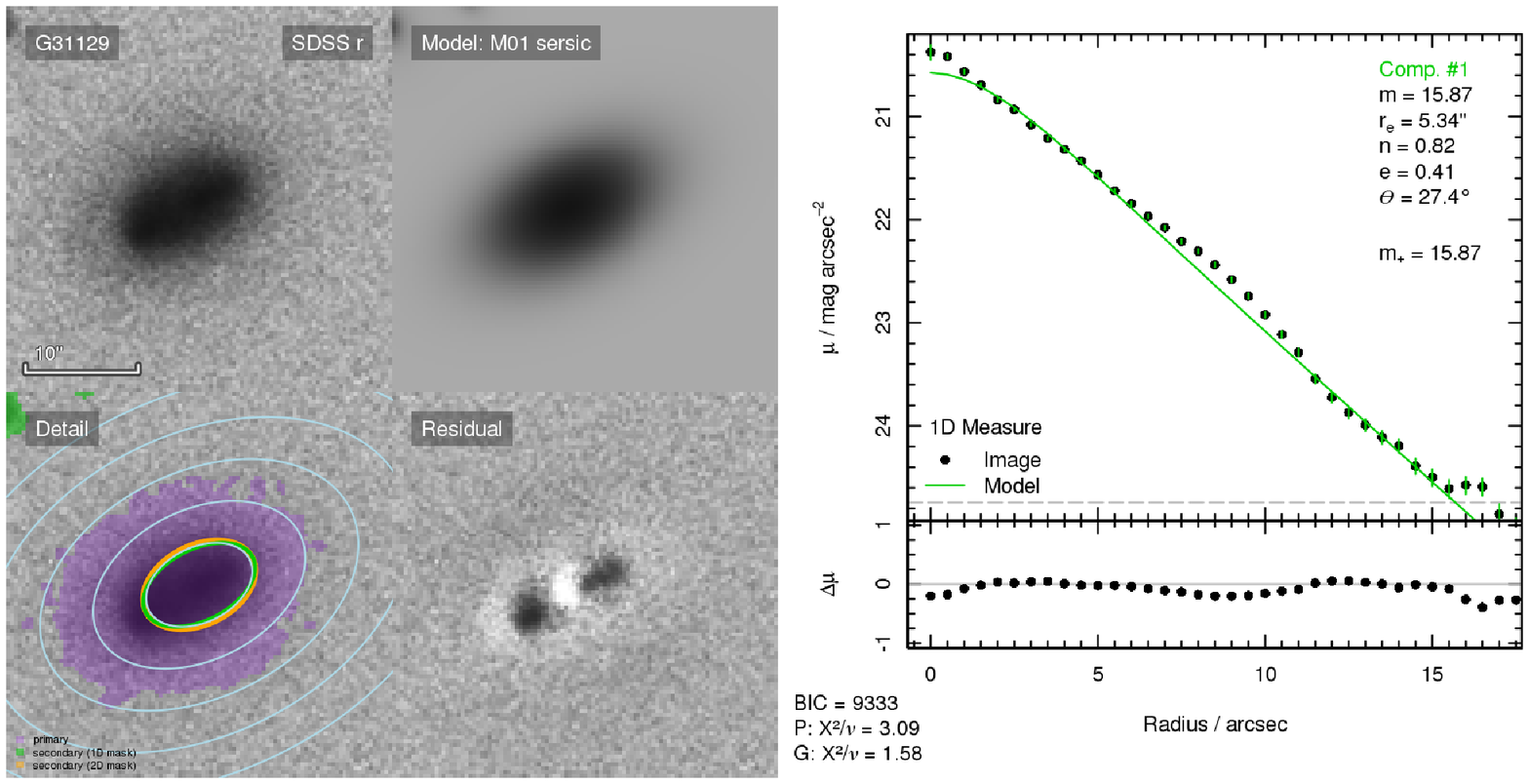,width=16cm}}}
{\rotatebox{0}{\epsfig{file=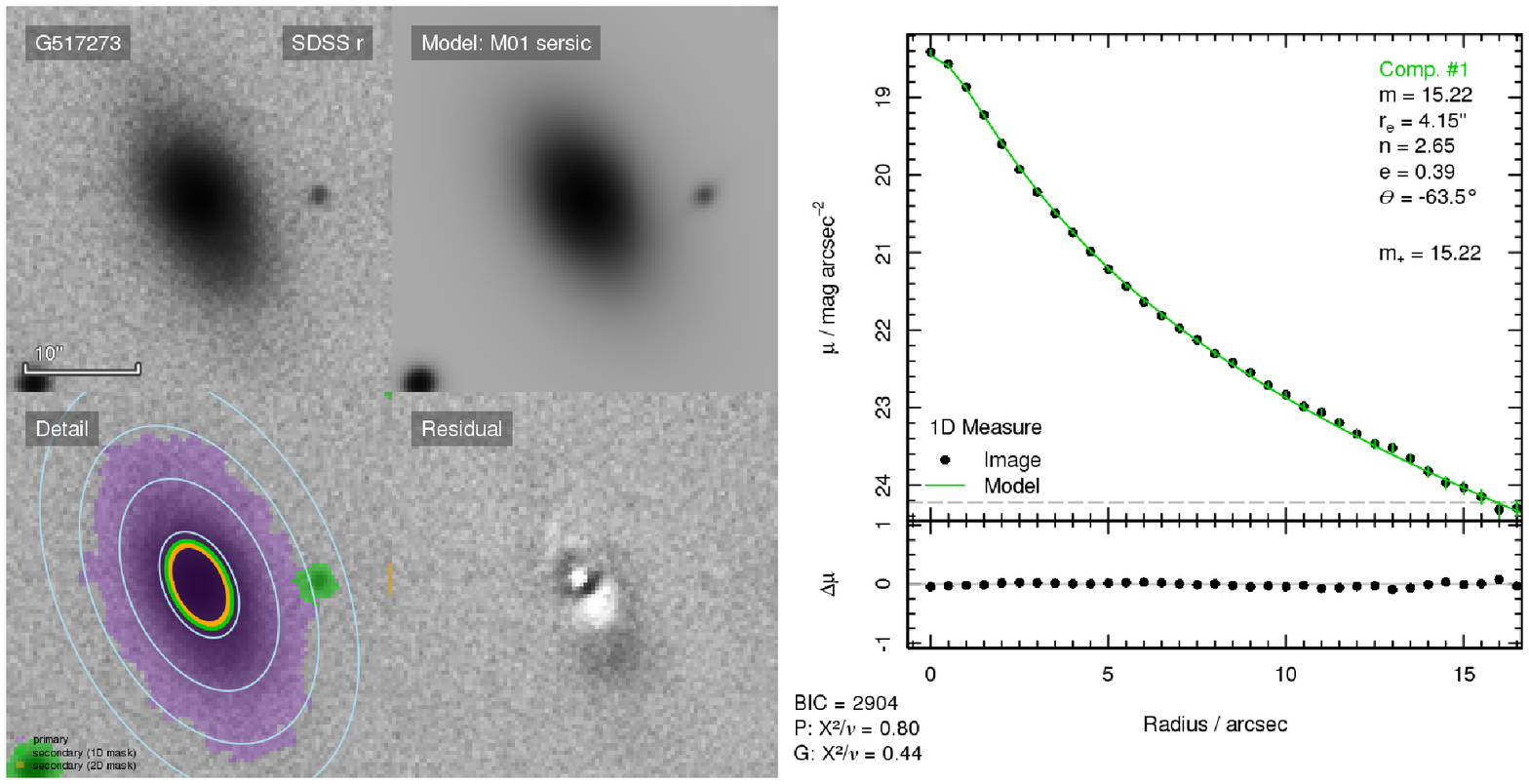,width=16cm}}}
\caption{The surface brightness fits in the {SDSS} $r$-band for two BSph galaxies. Each panel shows {\it (clockwise from top left:)}
 the original {SDSS} $r$-band image, S\'ersic model, 1D light profile (with residuals: image-model at the bottom), residual image and,
 ellipses centred on the primary galaxy used for estimating the light profile along with masked objects, respectively. }
\label{sigma}}
\end{figure*}

 \begin{figure*}
\centering{
{\rotatebox{0}{\epsfig{file=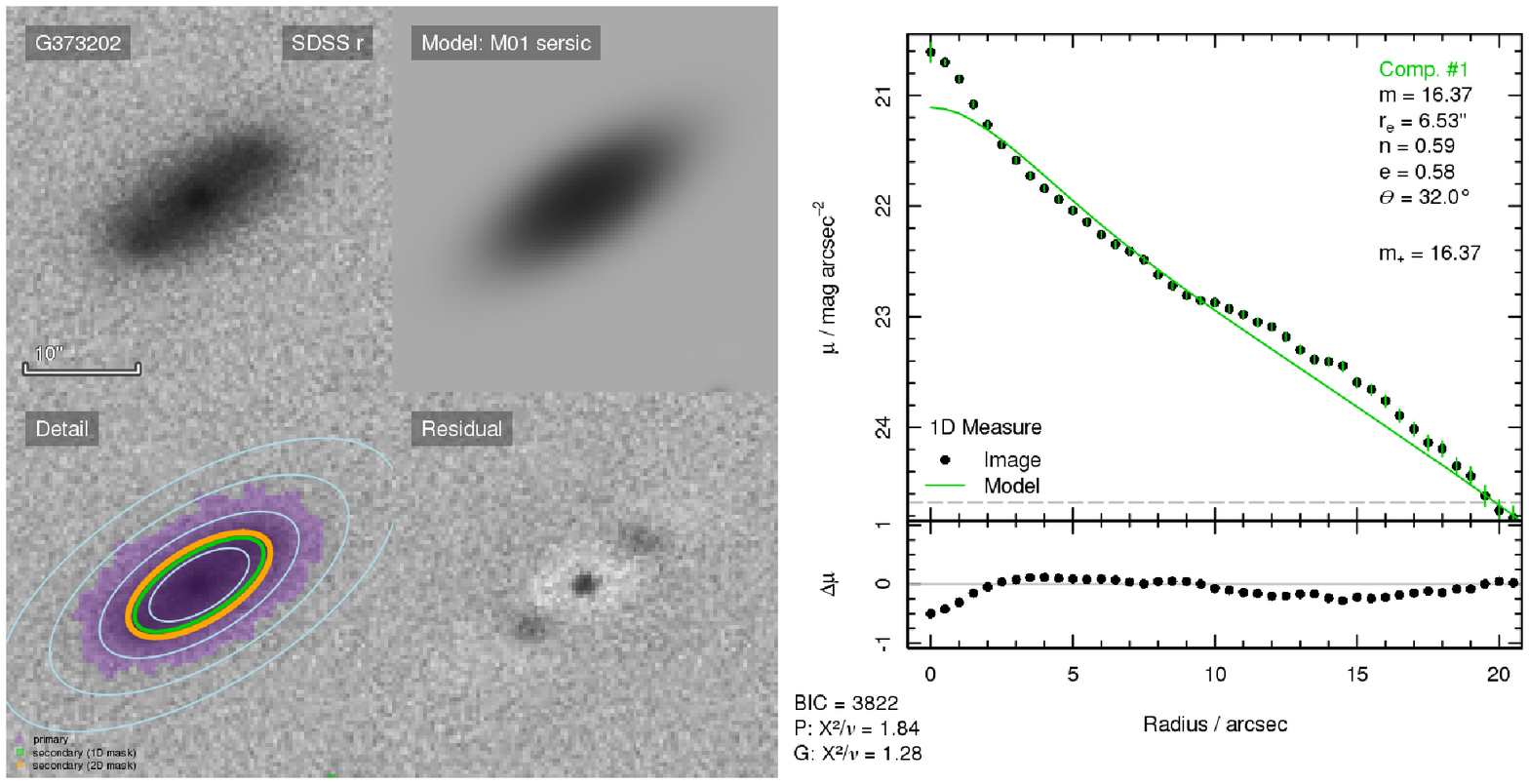,width=16cm}}}
{\rotatebox{0}{\epsfig{file=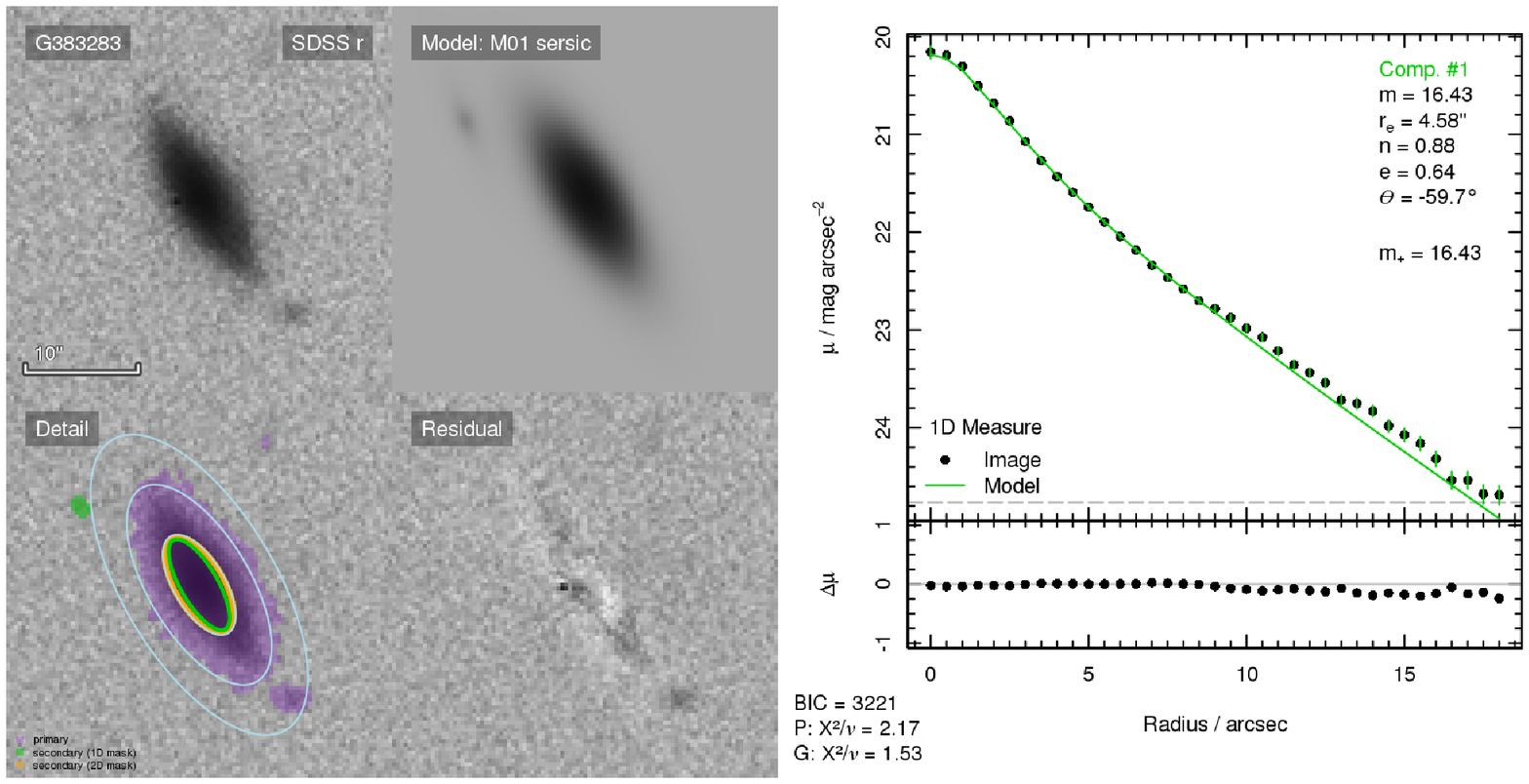,width=16cm}}}
\caption{Same as Figure~\ref{sigma}, but for elliptical galaxies.}
\label{sigma1}}
\end{figure*}

  \subsection{Structural properties}
 
 Light profile of all galaxies in the \g~database in 20 wavebands were fitted using a single \ser profile using the Structural Investigation of
 Galaxies via Model Analysis \citep[{\sc sigma}; v1.0-2][]{kelvin12}. The properties of our sample as outputted by \s~are described in
 Paper I. Figure~\ref{sigma} and \ref{sigma1} show two examples each of \bs~galaxies and ellipticals in SDSS $r$-band. The residual images
 evidently show the presence of a disk or a nuclear component in addition to the bulge modelled by 
 the single-\ser profile  in these systems. Such non-negligible residuals are found for $\sim 43$ per cent of ellipticals and 
 $\sim 38$ per cent of \bs~galaxies in our sample. This is consistent with the fraction ($\sim 42$ per cent) found by
 \citet{george17} using the SDSS Stripe 82 data for \bs~galaxies. Together these results elucidate that although this class of galaxies
 is termed \bs s here (and ``little blue spheroids"  or ``blue ellipticals" or ``blue early-type galaxies" elsewhere), structurally 
 they can be further subdivided into at least two different types: genuine blue spheroids, and blue dwarf elliptical galaxies which can
 be resolved into two components, and therefore should instead be called blue dwarf ``lenticulars" instead of ellipticals or spheroids.
   
\begin{figure*}
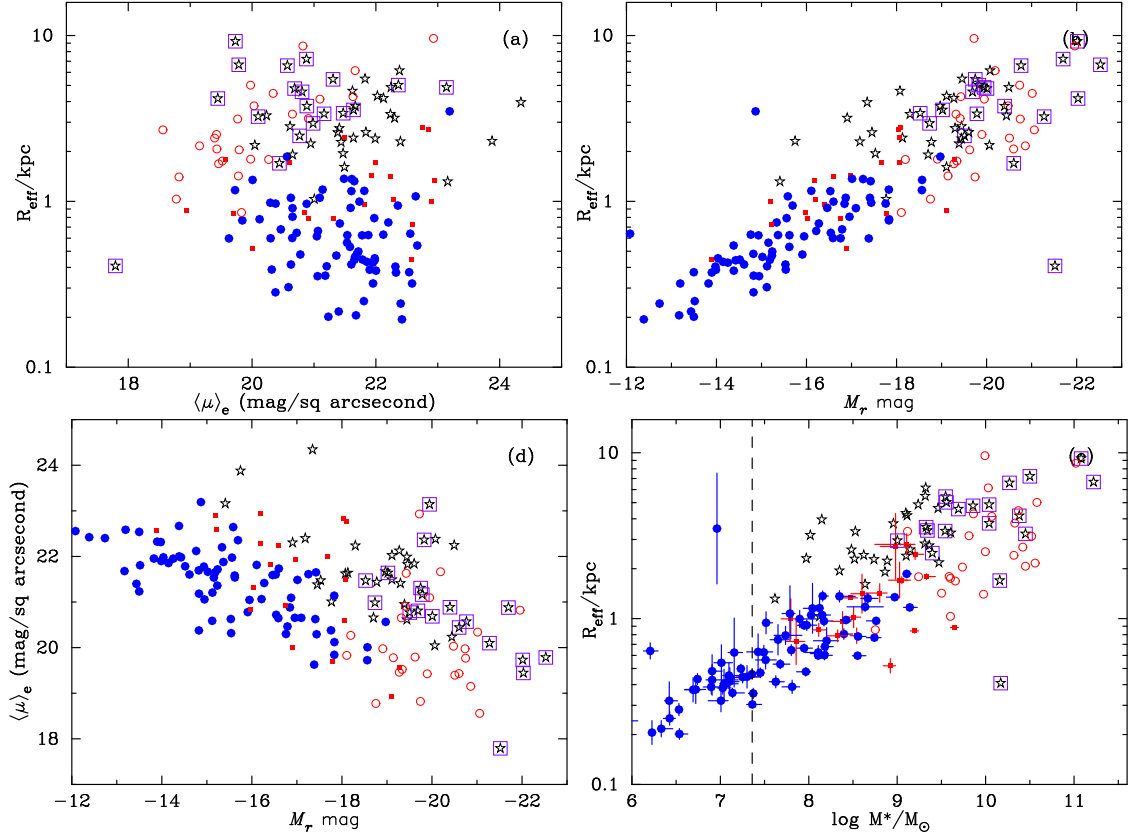

 \centering{
 \subfloat{{\rotatebox{270}{\epsfig{file=scaling-a.ps,width=5.5cm}}}}}
 \centering{
 \subfloat{{\rotatebox{270}{\epsfig{file=scaling-b.ps,width=5.5cm}}}}}
 \centering{
 \subfloat{{\rotatebox{270}{\epsfig{file=scaling-c.ps,width=5.5cm}}}}}
 \centering{
\subfloat{{\rotatebox{270}{\epsfig{file=scaling-d.ps,width=5.5cm}}}}}
 \caption{Scaling relations for bulge-dominated and spiral galaxies. Effective radius (\re) is shown as a function of
 (a) average surface brightness, (b) absolute $r$-band magnitude and (c) stellar mass. Panel (d) shows the 
 average surface brightness as a function of $M_r$.   }
 \label{radius}
 \end{figure*}

 In Figure~\ref{radius} we show the 2-d space mapped by the structural parameters \re~and average surface brightness
 ($\langle \mu\rangle_e$) as a function of $M_r$ and $M^*$ and the latter two with each other. Figure~\ref{radius} reiterates
 our results from Paper I, i.e. \bs~galaxies are structurally similar to ellipticals, spanning the same range of \re,
 $\langle \mu\rangle_e$, luminosity and $M^*$. Figure~\ref{radius} (c) shows that \bs s, ellipticals and red spirals
 form a single sequence in the stellar mass-\re~space, but star-forming spirals and lenticulars deviate away from the mean sequence 
 such that at any $M^*$ on average star-forming spirals are larger and lenticulars are smaller than the red spirals. 
 Figure~\ref{radius} (d) shows that all spheroids follow the same mean relation of declining $\langle \mu\rangle_e$ with increasing
 luminosity. But at fixed luminosity, spirals always have higher surface brightness relative to the spheroids.  
 
 In a nutshell this section shows that \bs~galaxies are structurally similar to their passively-evolving counterparts, but resemble
 star-forming spirals in age, $Z$ and star formation properties. In the following section we examine the clustering properties of 
 galaxies using automatic clustering algorithm and re-analyse their distribution in some of the parameter spaces discussed
 above.
   
\section{k-means clustering analysis}
\label{s:kmeans}

\begin{table*}
\caption{Automated classification results.}
\label{clusterlist}
\vspace*{3mm}
\begin{tabular}{cllrrrrrr}
\hline
 analysis & parameters   & $N_g$ & $N_c$ & $n_1$& $n_2$& $n_3$ & $p_{bsph}$ & $p_{other}$  \\
\hline
1 & age, Z, $M_*$           &  165  & 2 &   88  &  77 &    --  &   92\%&   24\%         \\
2 & age, Z, \ssf               & 165  & 2  &  120 & 45  &    --  &   97\% &  41\%        \\
3 & age, Z, \ssf, $M_{dust}$ & 165 & 3 & 74 & 50 & 41 & 85\%  & 16\%            \\ 
4 & age, Z, \ssf, $M_{dust}$, $M_*$ & 165 & 3 & 69 & 59 & 37 & 84\% & 12\% \\
\hline
\end{tabular}
\vspace*{3mm}

Note: $N_g$ is the number of galaxies in the sample, $N_c$ is the preferred
 number of clusters, and $n_i$ are the number of galaxies assigned to cluster $i$. $p_{bsph}$ is the percentage of
 all visually-classified \bs~galaxies assigned to the first cluster and $p_{other}$ is the percentage of all galaxies in the
 first cluster which were not classified as \bs.
\end{table*}

 \begin{figure}
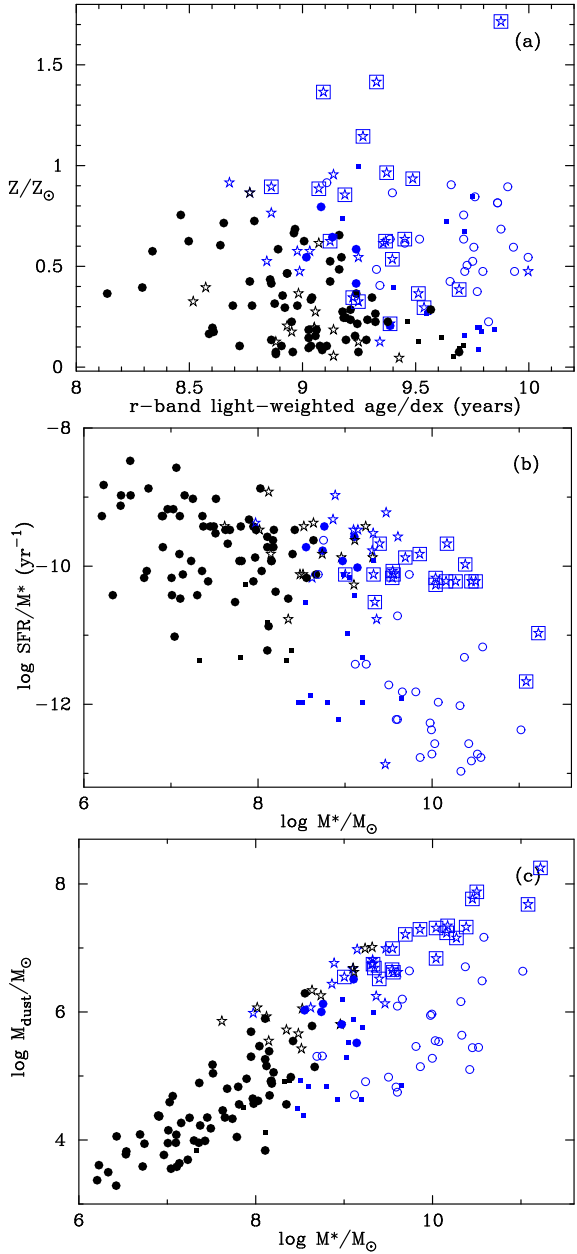

 \centering{
 {\rotatebox{270}{\epsfig{file=kmeans1a.ps,width=5.5cm}}}}
 \centering{
 {\rotatebox{270}{\epsfig{file=kmeans1b.ps,width=5.5cm}}}}
  \centering{
 {\rotatebox{270}{\epsfig{file=kmeans1c.ps,width=5.5cm}}}}
  \caption{Automated classification of galaxies in our sample. Panel (a) is the same as Figure~\ref{age-z}, 
 but colour-coded for ``clusters'' identified by the k-means algorithm using age, $Z$ and SFR/$M^*$. The other two panels 
 show the distribution of (b) specific SFR, and (c) $M_{dust}$ as a function of $M^*$ for the galaxies belonging to the
 two clusters preferred by k-means. The symbol types are same as in the above figures. Two clusters are preferred 
 statistically for this set of parameters. The \bs s are statistically well separated from the spirals as well as the ellipticals.   } 
 \label{kmeans1}
 \end{figure}

 \begin{figure}
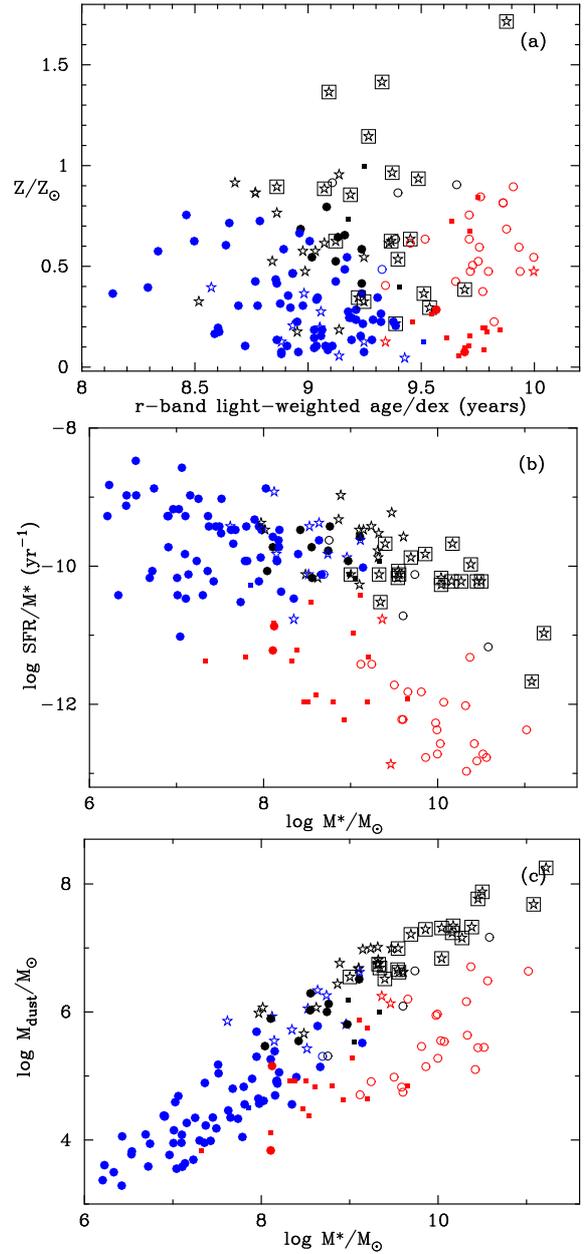

 \centering{
 {\rotatebox{270}{\epsfig{file=kmeans2a.ps,width=5.5cm}}}}
 \centering{
 {\rotatebox{270}{\epsfig{file=kmeans2b.ps,width=5.5cm}}}}
  \centering{
 {\rotatebox{270}{\epsfig{file=kmeans2c.ps,width=5.5cm}}}}
 \caption{Same as Fig.~\ref{kmeans1} but for parameters age, $Z$, SFR/$M^*$ and $M_{dust}$. In this case, 
 three clusters are preferred by k-means.  The \bs s are statistically well separated from the spirals as well the as ellipticals.  } 
 \label{kmeans2}
 \end{figure}

 One approach to test for connections between different galaxy populations is to examine the clustering properties of their parameters. Specifically, we used an objective clustering algorithm to test if the blue spheroidal galaxies were assigned to a group of their own, or if they clustered with either the spiral or the elliptical galaxies in our sample. We used the ``k-means'' algorithm \citep{macqueen67} to decompose the data into a specified number of clusters. For a given number of clusters, this finds the cluster positions that minimise the sum of the squares of the distances from each data point to its cluster centre. We determined the best number of clusters to adopt by using the NbClust \citep{charrad13} package in the R programming language. This package uses 23 different methods for determining the best number of clusters and selects the number proposed by the most methods. Before starting the clustering analysis we removed any objects with missing data and then scaled the remaining galaxies to have a mean of zero and standard deviation of unity in each parameter. All the parameters we analysed were logarithmic measurements (or magnitudes). This approach is very similar to our analysis in Paper 1.

 We applied the clustering analysis to several combinations of MAGPHYS parameters reflecting the stellar populations: age, metallicity, stellar mass,
 dust mass and specific star formation rate. These are listed in Table~\ref{clusterlist}. 

For the first analysis, we considered age, metallicity and $M^*$. In this case, two clusters were preferred, containing 88 and 77 galaxies respectively (see Table \ref{clusterlist}). The larger cluster has galaxies at lower ages, metallicities, stellar masses and dust masses, as shown in Figs.~\ref{kmeans1} (a), (b) and (c). This cluster contains 92 per cent of the blue spheroids, but is strongly (24 per cent) contaminated by other galaxy types. The smaller cluster contains 67 per cent of the spiral galaxies and 87 per cent of the passive (elliptical and lenticular) galaxies. This partition has therefore demonstrated some separation between the \bs s and both the other main galaxy types.

For the second analysis, we replaced stellar mass with specific star formation. In this case also two clusters were preferred (see Table \ref{clusterlist}). The largest cluster again contained galaxies at lower ages, metallicities, dust masses and stellar masses. This cluster contains 97 per cent of the \bs s, but is heavily (41 per cent) contaminated by other galaxy types than for the first analysis.   

 For the third analysis, we added in dust mass. In this case three clusters were preferred, as shown in Figs.~\ref{kmeans2}
  (a), (b) and (c) (see Table \ref{clusterlist}).The largest cluster 
 comprising \bs s contained 74 galaxies and 85 per cent of all \bs s in our sample, but is comparatively less contaminated (16 per cent)
 by other galaxy types. We further note that adding in stellar mass to this list (analysis 4) reduces the contamination by other galaxy types, 
 by effectively removing some of the spirals from the cluster containing \bs s, but does not increase the completeness fraction
 of \bs s in their cluster.
 
 In summary therefore, the clustering analyses we applied have separated the \bs s from both the spiral and the
 passively-evolving spheroids in our sample. This also evidently shows that once all the four parameters namely, age, $Z$, SFR/$M^*$
 and $M_{dust}$ are taken into account, \bs s are statistically different from both spirals and ellipticals.


 \section{Atomic gas mass}
 \label{s:h1}
 
 \begin{figure}
 \centering{
 {\rotatebox{270}{\epsfig{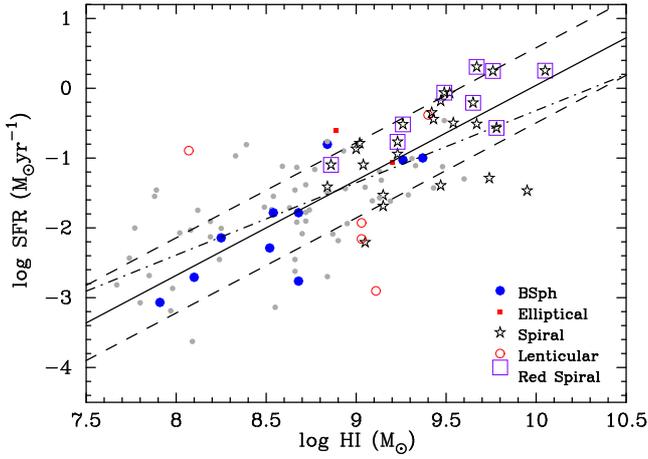}}}}
 \caption{SFR as a function of the atomic gas mass for the galaxies in our sample for which HI data are available. 
 The coloured symbols are the same as in Fig.~\ref{mag-mass}, while the grey points represent irregular and low surface
 brightness galaxies. The solid and dashed lines are the linear least squares fit and $\pm 1 \sigma$ deviation in it,
 respectively considering 10 BSphs and 26 spiral galaxies only. The dot-dashed line represents the least squares fit
 line for all the 101 galaxies detected in HI by the ALFALFA survey.  }
 \label{h1}
 \end{figure}

 Recent and current evolution of star formation properties of galaxies are strongly dependent on the amount of 
 gas contained in them. Similarities in the star formation properties of \bs~galaxies and spirals suggests that they may obey 
 similar SFR-$M_{gas}$ scaling relation. In order to test this hypothesis we make use of the Arecibo Legacy Fast ALFA 
 survey \citep[ALFALFA;][]{giovanelli05} which is a blind extragalactic HI survey done using the Arecibo telescope to
 conduct a census of the local HI universe over a cosmologically significant volume.

 101/428 galaxies in our sample have been detected by the ALFALFA survey with median(mean) signal-to-noise ratio (SNR) of
 12(25), and $\rm{SNR} \geq 4.6$. To find the HI counterparts, we matched the LAMBDAR position
 coordinates to the coordinates of the optical counterpart found by the ALFALFA team \citep{haynes11}. 
  All except 7 of the 101 sources from our sample are matched to counterparts within $5^{\prime\prime}$ 
 (maximum separation 9$^{\prime\prime}$) from the optical counterpart of the HI source as described above; 
 five of these are irregular galaxies, and one each is an LSB and a spiral galaxy, respectively. These 7 sources are
 included in Figures~\ref{h1} and \ref{h1-ratio}.
 
 In Figure~\ref{h1} we show the SFR of different morphological types of galaxies as a function of their HI mass. For completeness,
 we show all the 101 galaxies from our sample detected by the ALFALFA survey. For all galaxies,  
 SFR is positively correlated with HI mass with a slope of $1.36 \pm 0.54$.
 The red spirals lie above the mean relation, suggesting that these galaxies have higher SFR than expected for their HI mass.
 The BSphs seem to follow the mean scaling relation as all the other star-forming galaxies in this space. These
 observations suggest that even if a \bs~develops a disk and moves rightward towards higher $M^*$ and
 SFR in Figure~\ref{h1}, it is likely to obey the same scaling relation as the spirals, only a few of which significantly
 deviate away from it. But since only two ellipticals are detected in HI, these
 observations are inconclusive about the fate of the \bs s once the star formation and gas supply fades away.


 \section{Discussion}
 \label{discuss}
 
 The aim of this paper is to test whether BSph galaxies could be progenitors of spirals or passively-evolving
 elliptical galaxies or if they form a different class of their own based on their structural, physical and star formation properties.
  \bs~galaxies form around 20 per cent of all low-mass
 (\smass $< 10^9$) galaxies, and 82.5 per cent of all low-mass spheroids in our sample, which is in agreement with the 
 fractions found by \citet{k09}. Furthermore, \citet{moffett16} found that  the ``little blue spheroids" classified in the GAMA
 visual morphology DMU contribute $\sim 2\%$ of the stellar mass density at $0.002<z<0.06$. 
 Our analysis showed that \bs s are structurally very similar to low-mass elliptical galaxies. The fraction of elliptical
 and \bs~galaxies which are likely to contain disk or other components missed by a single \ser profile used for fitting
  the galaxies is also similar \citep[Figs.~\ref{sigma} and \ref{sigma1}; also see][]{george17}. 
  
  The deep imaging data from the SDSS Stripe 82 has
 been examined to look for tidal features in the red sequence elliptical galaxies and BSphs. While \citet{kaviraj10} 
 found that only 28$\pm$3 per cent of the 238 ellipticals in their sample show tidal features, \citet{george17} found a fraction
 of 58$\pm$7 per cent for the BSphs using analysis of residual images similar to the ones shown in Fig.~\ref{sigma}. 
 \citet{george17} also found that tidal features are more common around massive BSphs (\smass $> 10^{10.5}$), which 
 are absent in our sample. But since the detection of such features depends upon the depth of the
 imaging data \citep[see for instance table 3 of][]{kim12}, it is not surprising that we fail to find any significant differences
 between the residual images of ellipticals and BSphs using the shallow SDSS $r$-band imaging data.
 
\begin{figure}
 \centering{
 {\rotatebox{270}{\epsfig{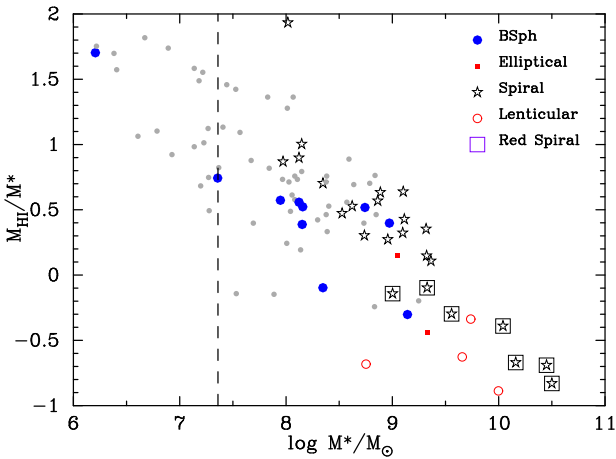}}}}
 \caption{This figure shows the ratio of HI and stellar mass of galaxies detected by the ALFALFA as a function of their
 stellar mass. The dashed line is the limiting stellar mass for our sample as per Figure~\ref{mag-mass}.  }
 \label{h1-ratio}
 \end{figure}

 The scaling relations (Figure~\ref{radius}) for spheroids and spirals shows that BSphs occupy similar range of parameters:
 \re, $M^*$ and \mue~as the red ellipticals. This implies that if BSphs are ellipticals experiencing strong starburst, the 
 phenomenon causing starburst has left their morphology unchanged. The 10 BSphs from our sample which were detected
 by the ALFALFA survey show a significant amount of atomic gas. Figure~\ref{h1-ratio} shows that the median (and mean) ratio of
atomic-gas-to-stellar-mass in BSphs is $\sim 0.5$. It also shows that even in this small sample of BSphs the ratio 
 $M_{HI}/M^*$ varies by $\sim 2$ dex over an $M^*$ range of two orders of magnitude, suggesting that the evolution
  history of BSphs is very heterogenous.   
 
  \begin{figure}
 \centering{
 {\rotatebox{270}{\epsfig{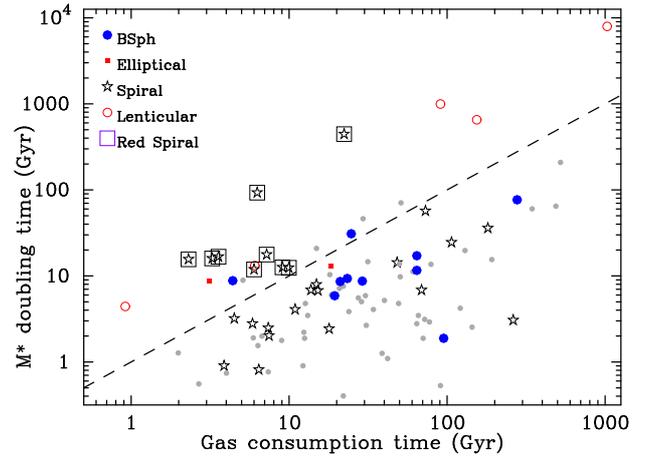}}}}
 \caption{The stellar mass doubling time as a function of the gas depletion time scale for the
 ALFALFA detected galaxies in our sample. The dashed line represents the path followed by galaxies which have enough gas to
 double their stellar mass by continuing to form stars at the present rate. Symbols are same as in Figure~\ref{h1}. }
 \label{time}
 \end{figure}

 Following \citet{k09}, we estimated the stellar mass doubling time as the stellar mass divided by the SFR, and
 gas depletion time scale as the atomic gas mass divided by the SFR. 
 These time scales are crude approximations uncorrected for future infall of new gas or decline of star formation.
 Figure~\ref{time} shows these two time scales for the galaxies in our sample for which HI data are available.
 Assuming that the unmeasured molecular gas mass in these galaxies is not very
 large, galaxies on the left of the line of equality can not double their stellar mass without accreting new gas.
 Figure~\ref{time} therefore shows that most \bs~galaxies (8/10) can evolve significantly even under the
 hypothetical closed-box scenario assumed here, and hence may not undergo any morphological changes. 
 
 But if a \bs~needs to develop a disk similar to a spiral galaxy its radius must increase by a factor of $\sim 3$ (Figure~\ref{radius}),
 implying a large increase in mass. So along with the presence of large amount of gas, \bs s should form stars with 
 great efficiency. The gas consumption time-scale shown in Figure~\ref{time} crudely represents the inverse
  of star formation efficiency, i.e. $SFE=SFR/M_{HI}$. Even in our small sample it is evident that the \bs s are on average less
 efficient at forming stars than spiral galaxies. It is therefore plausible but not mandatory that at least some of the
 \bs~galaxies may evolve into small disk galaxies \citep[also see][]{noeske06}.
 
  \begin{figure}
 \centering{
 \subfloat{{\rotatebox{270}{\epsfig{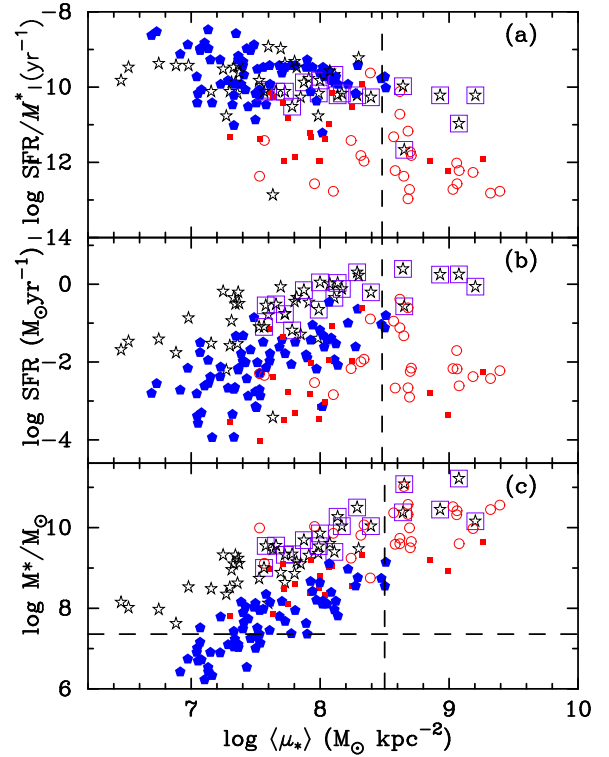}}}}}
  \caption{The stellar surface mass density $\mu_*$ at \re~as a function of (a) \ssf, (b) SFR and
 (c) $M^*$ for the spheroids and spirals in our sample. The vertical dashed line is the characteristic stellar surface mass
 density of $3\times10^8 M_{\odot}$ kpc$^{-2}$ reported by \citet{kauff06}. }
 \label{smd}
 \end{figure}

 Using nearby ($z\sim0.1$) galaxies more massive than a few times $10^9 M_{\odot}$ from the SDSS, \citet{kauff06} reported
 a critical stellar surface mass density, log $\mu_* \sim 3\times10^8 M_{\odot}$ kpc$^{-2}$ above which the star
 formation in disk-dominated galaxies occurs in short-lived intense bursts. We estimated $\mu_* = M^*/\pi R_{\rm{eff}}^2$
 for all the galaxies in our sample. Figure~\ref{smd} shows $\mu_*$ for the spheroids
 and spiral galaxies\footnote{Two \bs s and one red spiral galaxy at $\mu_* =$ 5.37, 6.09 and 10.45  $M_{\odot}$ kpc$^{-2}$ 
do not appear in the figure because of the chosen range for the ordinate. Both the \bs s fall below the mass completeness limit;
 log $M^*=10.17 M_{\odot}$ for the red spiral.} 
 as a function of their $M^*$, SFR and sSSFR, respectively. For our sample only 28 galaxies have $\mu_*$
 greater than the critical threshold, of which 17 (61 per cent) are passively-evolving lenticulars and six (21 per cent) are red spirals.  
 
 Figure~\ref{smd} shows that the stellar surface mass density is strongly correlated with stellar mass irrespective of
 galaxy morphology. The SFR is found to be a function of $\mu_*$ for the \bs s and spirals, but not for the ellipticals and lenticulars.
 The sSFR for \bs~and spiral galaxies is independent of $\mu_*$ albeit with large scatter, some of which may be attributed to
 the weak correlation between $M^*$ and SFR (Figure~\ref{sfr}). The sSFR for red spirals is remarkably constrained
 to $\sim 10^{-10}$ yr$^{-1}$ despite their $M^*$ and $\mu_*$ varying by as much as two dex. For their relatively more
 massive sample, \citet{kauff06} found that sSFR remains constant for all galaxies with log $\mu_* < 8.5$ $M_{\odot}$
  kpc$^{-2}$, unlike our sample
 where low-mass ellipticals and passively-evolving lenticulars even below the critical $\mu_*$ have lower sSFR than 
 their star-forming counterparts. 
 
 We can conclude two things from Figure~\ref{smd}, firstly the criticality of the $\mu_*$ threshold below which the 
 mean sSFR of galaxies of a given mass and stellar surface density does not not depend upon either mass or surface density
 \citep{kauff06}, is invalid for our sample with a lower average stellar mass than the sample of Kauffmann et al.. Secondly, the \bs s
 follow a linear relation similar to spirals in the $\mu_*$-$M^*$ space, but with a different normalization factor. This analysis 
 therefore shows that for our sample, $M^*$ is a function of surface density irrespective of galaxy morphology, unlike the 
  SFR and sSFR for which the trends vary with galaxy morphology.
   
 \section{Summary}
 \label{summary}
 
 In this paper we examine the likelihood of \bs~galaxies as progenitors of spiral or elliptical galaxies.
 In order to do so, we make use of the data products derived from the panchromatic imaging and optical spectra of
 the GAMA galaxies. This paper presents our complete sample of 432 galaxies (also used in Paper I)
 spanning the redshift range $0.002<z<0.02$, and devoid of very high-density environments.

 We find that even though \bs s are structurally very similar to ellipticals, the distribution of their luminosity-weighted 
 age, $Z$, $M_{dust}$ and sSFR are more like the star-forming spirals than the passively-evolving spheroids (ellipticals
 or lenticulars). \bs s also follow the
 same SFR-$M_{HI}$ and $M_{HI}$-$M^*$ scaling relation as the spirals. We further showed that at any given $M^*$, \bs s are
 more compact than spirals and on average have higher stellar surface mass density at \re~than spiral galaxies, implying
 that on average their star formation efficiency is lower than spiral galaxies. 
  
 The automated clustering algorithm k-means applied to the multi-dimensional parameter space mapped by
 age, $Z$ and $M^*$ decomposes the spheroids and spiral galaxies into two ``clusters".
 The larger of these comprise 92 per cent of the \bs s but is heavily contaminated (24 per cent) by other types
 of galaxies having low age, $Z$ and $M^*$.
 Adding in the $M_{dust}$ to the multi-dimensional space yields a preference for three clusters, of which the one containing
 low age, $Z$, $M_{dust}$ and $M^*$ and high sSFR galaxies comprises 85 per cent of the \bs s but is less contaminated (16 per cent)
 by other galaxy types.
  
  The gas supply for star formation in a galaxy is regulated by the infall of new material onto the parent dark matter halo as it grows.
  In massive halos at later times, the time required by the gas to fall and cool becomes much larger than the dynamical time and
  therefore halts the supply of cold gas for the galaxy. For $\Lambda$CDM cosmology, it has been shown that the transition between
  infall-regulated and cooling flow regime occurs at a halo mass of around $10^{12}$ M$_\odot$ \citep{birnboim03,keres05}.
  
  Most of the galaxies in our sample, particularly all the low-mass galaxies must have infall-regulated supply of gas.  
  Their future is therefore likely to depend on environment: \bs s in the low-density region can accrete
  more cold gas, therefore developing an intermediate-- or large--scale disk, while those in the high density environment 
  are prone to more lumpy accretion building an elliptical galaxy. A bit of both the processes is likely to result in a
  low-mass lenticular. It is possible that this is an ongoing process at all redshifts but because these galaxies have
  low mass, we are only able to observe them locally.

 To conclude, our data suggest that although \bs~galaxies are structurally similar to ellipticals and have physical and 
 star-formation properties like spirals, statistically they are distinguishable from either of them in the multi-dimensional
 parameter space mapped by age, $Z$, $M_{dust}$ and sSFR. Therefore, based on our analysis we conclude that 
 some \bs s may evolve into disk galaxies in the future, while others in the low-density environments may evolve into
 small red ellipticals. But their currently observable properties statistically distinguish them from spirals as well as ellipticals.
 
 \section{Acknowledgements}
 
 GAMA is a joint European-Australasian project based around a spectroscopic campaign using the Anglo-Australian
 Telescope. The GAMA input catalogue is based on data taken from the Sloan Digital Sky Survey and the UKIRT Infrared
 Deep Sky Survey. Complementary imaging of the GAMA regions is being obtained by a number of independent
 survey programmes including GALEX MIS, VST KiDS, VISTA VIKING, WISE, Herschel-ATLAS, GMRT and ASKAP
 providing UV to radio coverage. GAMA is funded by the STFC (UK), the ARC (Australia), the AAO, and the
 participating institutions. The GAMA website is http://www.gama-survey.org/ .  We are grateful to the reviewer for their
 suggestions and comments which helped improve this manuscript.

 Mahajan is funded by the INSPIRE Faculty award (DST/INSPIRE/04/2015/002311), Department of Science and
 Technology (DST), Government of India.
 
  \label{lastpage}


\label{lastpage}


\begin{thebibliography}{99}

 \bibitem[\protect\citeauthoryear{Abazajian et al.}{2009}]{abazajian09} Abazajian K.~N., et al., 2009, ApJS, 182, 543 

\bibitem[\protect\citeauthoryear{Baldry et al.}{2010}]{baldry10} Baldry I.~K., et al., 2010, MNRAS, 404, 86 
\bibitem[\protect\citeauthoryear{Birnboim \& Dekel}{2003}]{birnboim03} Birnboim Y., Dekel A., 2003, MNRAS, 345, 349 

 \bibitem[\protect\citeauthoryear{Cameron et al.}{2009}]{cameron09} Cameron E., Driver S.~P., Graham A.~W., Liske J.,
 2009, ApJ, 699, 105 
\bibitem[\protect\citeauthoryear{Charrad et al.}{2012}]{charrad13} Charrad M., Ghazzali N., Boiteau V., Niknafs A., 2012, 
 UseR! 2012, CGB12a

 \bibitem[\protect\citeauthoryear{da Cunha, Charlot, \& Elbaz}{2008}]{dacunha08} da Cunha E., Charlot S., Elbaz D.,
  2008, MNRAS, 388, 1595 

 \bibitem[\protect\citeauthoryear{Davies et al.}{2016}]{davies16} Davies L.~J.~M., et al., 2016, MNRAS, 455, 4013 
 \bibitem[\protect\citeauthoryear{Driver et al.}{2006}]{driver06} Driver S.~P., et al., 2006, MNRAS, 368, 414 
  \bibitem[\protect\citeauthoryear{Driver et al.}{2011}]{driver11} Driver S.~P., et al., 2011, MNRAS, 413, 971 
 \bibitem[\protect\citeauthoryear{Driver et al.}{2016}]{driver16} Driver S.~P., et al., 2016, MNRAS, 455, 3911 

\bibitem[\protect\citeauthoryear{Gordon et al.}{2017}]{gordon17} Gordon Y.~A., et al., 2017, MNRAS, 465, 2671
 \bibitem[\protect\citeauthoryear{Graham, Dullo, \& Savorgnan}{2015}]{graham15} Graham A.~W., Dullo B.~T.,
 Savorgnan G.~A.~D., 2015, ApJ, 804, 32 
 \bibitem[\protect\citeauthoryear{Graham, Ciambur, \& Savorgnan}{2016}]{graham16} Graham A.~W., Ciambur B.~C.,
  Savorgnan G.~A.~D., 2016, ApJ, 831, 132 
 \bibitem[\protect\citeauthoryear{Graham et al.}{2017}]{graham17} Graham A.~W., Janz J., Penny S.~J.,
  Chilingarian I.~V., Ciambur B.~C., Forbes D.~A., Davies R.~L., 2017, ApJ, 840, 68 

 \bibitem[\protect\citeauthoryear{George}{2017}]{george17} George K., 2017, A\&A, 598, A45 
 \bibitem[\protect\citeauthoryear{Giovanelli et al.}{2005}]{giovanelli05} Giovanelli R., et al., 2005, AJ, 130, 2598 
\bibitem[\protect\citeauthoryear{Gunawardhana et al.}{2013}]{gunawardhana13} Gunawardhana M.~L.~P., et al., 2013,
 MNRAS, 433, 2764 

\bibitem[\protect\citeauthoryear{Haynes et al.}{2011}]{haynes11} Haynes M.~P., et al., 2011, AJ, 142, 170 
\bibitem[\protect\citeauthoryear{Hopkins et al.}{2003}]{hopkins03} Hopkins A.~M., et al., 2003, ApJ, 599, 971 
\bibitem[\protect\citeauthoryear{Hopkins et al.}{2013}]{hopkins13} Hopkins A.~M., et al., 2013, MNRAS, 430, 2047 

\bibitem[\protect\citeauthoryear{Janowiecki et al.}{2017}]{janowiecki17} Janowiecki S., Catinella B., Cortese L.,
 Saintonge A., Brown T., Wang J., 2017, MNRAS, 466, 4795 

\bibitem[\protect\citeauthoryear{Kannappan}{2004}]{k04} Kannappan S.~J., 2004, ApJ, 611, L89 
\bibitem[\protect\citeauthoryear{Kannappan \& Wei}{2008}]{k08} Kannappan S.~J., Wei L.~H., 2008, AIPC, 1035, 163 
\bibitem[\protect\citeauthoryear{Kannappan, Guie, \& Baker}{2009}]{k09} Kannappan S.~J., Guie J.~M., Baker A.~J., 2009, AJ, 138, 579 
\bibitem[\protect\citeauthoryear{Kauffmann et al.}{2006}]{kauff06} Kauffmann G., Heckman T.~M., De Lucia G., Brinchmann J., Charlot S., Tremonti C., White S.~D.~M., Brinkmann J., 2006, MNRAS, 367, 1394 
\bibitem[\protect\citeauthoryear{Kaviraj}{2010}]{kaviraj10} Kaviraj S., 2010, MNRAS, 408, 170 
\bibitem[\protect\citeauthoryear{Kelvin et al.}{2012}]{kelvin12} Kelvin L.~S., et al., 2012, MNRAS, 421, 1007 
\bibitem[\protect\citeauthoryear{Kennicutt}{1998}]{kennicutt98} Kennicutt R.~C., Jr., 1998, ARA\&A, 36, 189 
\bibitem[\protect\citeauthoryear{Kere{\v s} et al.}{2005}]{keres05} Kere{\v s} D., Katz N., Weinberg D.~H., Dav{\'e} R.,
 2005, MNRAS, 363, 2 
\bibitem[\protect\citeauthoryear{Kim et al.}{2012}]{kim12} Kim T., et al., 2012, ApJ, 753, 43 

\bibitem[\protect\citeauthoryear{Liske et al.}{2003}]{liske03} Liske J., Lemon D.~J., Driver S.~P., Cross N.~J.~G.,
 Couch W.~J., 2003, MNRAS, 344, 307 
\bibitem[\protect\citeauthoryear{Liske et al.}{2015}]{liske15} Liske J., et al., 2015, MNRAS, 452, 2087 
\bibitem[\protect\citeauthoryear{Lopes et al.}{2016}]{lopes16} Lopes P.~A.~A., Rembold S.~B., Ribeiro A.~L.~B.,
 Nascimento R.~S., Vajgel B., 2016, MNRAS, 461, 2559 
 \bibitem[\protect\citeauthoryear{Loveday et al.}{2012}]{loveday12} Loveday J., et al., 2012, MNRAS, 420, 1239 
 
  \bibitem[\protect\citeauthoryear{MacQueen}{1967}]{macqueen67} Macqueen J., ``Some methods for classification and analysis of
  multivariate observations", Proc. Fifth Berkeley Symp. on Math. Statist. and Prob., Vol. 1 (Univ. of Calif. Press, 1967), 281, 297
  \bibitem[\protect\citeauthoryear{Moffett et al.}{2016}]{moffett16} Moffett A.~J., et al., 2016, MNRAS, 462, 4336 
 \bibitem[\protect\citeauthoryear{Mahajan et al.}{2015}]{mahajan15} Mahajan S., et al., 2015, MNRAS, 446, 2967 (Paper I)
 
 \bibitem[\protect\citeauthoryear{Noeske et al.}{2006}]{noeske06} Noeske K.~G., Koo D.~C., Phillips A.~C., Willmer C.~N.~A.,
 Melbourne J., Gil de Paz A., Papaderos P., 2006, ApJ, 640, L143 

 \bibitem[\protect\citeauthoryear{Robotham et al.}{2010}]{robotham10} Robotham A., et al., 2010, PASA, 27, 76 

\bibitem[\protect\citeauthoryear{Schawinski et al.}{2009}]{s09} Schawinski K., et al., 2009, MNRAS, 396, 818 

\bibitem[\protect\citeauthoryear{Tonry et al.}{2000}]{tonry00} Tonry J.~L., Blakeslee J.~P., Ajhar E.~A., Dressler A., 2000, ApJ, 530, 625 


 \bibitem[\protect\citeauthoryear{Wright et al.}{2016}]{wright16} Wright A.~H., et al., 2016, MNRAS, 460, 765 
 
\end{thebibliography}
\end{document}